\documentclass[12pt,a4paper]{article}
\usepackage{amsmath}
\usepackage{graphicx}
\usepackage{wrapfig}
\usepackage{amsfonts}
\usepackage{amssymb}

\begin{document}

\title{Acceleration of the Universe as a consequence of gravitation 
properties}
\author{L.V.Verozub and A.Y.Kochetov\\
Kharkov National University, Kharkov, Ukraine\\                       
email: verozub@gravit.kharkov.ua}
\date{}
\maketitle
\begin{abstract}                                                      
The analysis of the data from the distant supernovae  (A.Riess et al, 
Astron.J. {\bf 116}, 1009, (1988)) for acceleration of the expending
Universe from the viewpoint of the
gravitation equations proposed by one of the authors (Phys.Lett. {\bf 156}, 
404 (1991)) is given. It is shown that the result from the above data 
that the deceleration parameter $q_{0}$ is  negative is a natural 
consequence
of the property of the gravitation force which follows from the above 
gravitation 
equations . It is an alternative explanation to general 
relativity
where a nonzero cosmological constant is used to explain the data.
\end{abstract}

\section{Introduction.}
Thirring \cite{Thirring} proposed that gravitation can be described as a
tensor field $\psi_{\alpha\beta}(x)$ of spin two in 4-dimensional
Pseudo-Euclidean space-time $E_{4}$ where the Lagrangian action describing 
the
motion of test particles in a given field is of the form
\begin{equation}
L=-m_{p}c\left[  g_{\alpha\beta}(\psi)\dot{x}^{\alpha}\dot{x}^{\beta}
\right]^{1/2}\;.
\end{equation}

In this equation $g_{\alpha\beta}$ is a tensor function of 
$\psi_{\alpha\beta
}$, $m_{p}$ is the paticle mass, $c$ is the speed of light and $\dot
{x}^{\alpha}=dx^{\alpha}/dt$ .

A theory based on that action must be invariant under the gauge
transformations 
$\psi_{\alpha\beta}\longrightarrow\bar{\psi}_{\alpha\beta}$
that are a consequence of the existence of ''extra'' components of the 
tensor
$\psi_{\alpha\beta}$. The transformations $\psi_{\alpha\beta}
\longrightarrow
\bar{\psi}_{\alpha\beta}$ give rise to some transformations 
$g_{\alpha\beta}$
$\longrightarrow$ $\bar{g}_{\alpha\beta}$. Therefore, the field equations 
for
$g_{\alpha\beta}(x)$ and equations of the motion of the test particle must be
invariant under these transformations of the tensor $g_{\alpha\beta}$. A
theory that are invariant with respect to the arbitrary gauge 
transformations
was proposed in the paper \cite{Verozub1}. The gravitation equations are 
of
the form%

\begin{equation}
B_{\alpha\beta;\gamma}^{\gamma}-B_{\alpha\delta}^{\epsilon}
B_{\beta\epsilon
}^{\delta}=0 \label{myeqs}%
\end{equation}
where
\begin{equation}
B_{\alpha\beta}^{\gamma}=\Pi_{\alpha\beta}^{\gamma}-\overset{\circ}{\Pi
}_{\alpha\beta}^{\gamma}, \label{tensB}%
\end{equation}%
\begin{equation}
\Pi_{\alpha\beta}^{\gamma}=\Gamma_{\alpha\beta}^{\gamma}-(n+1)^{-1}\left[
\delta_{\alpha}^{\gamma}\Gamma_{\epsilon\beta}^{\epsilon}-\delta_{\beta
}^{\gamma}\Gamma_{\epsilon\alpha}^{\epsilon}\right]  , \label{Thomases}%
\end{equation}%

\begin{equation}
\overset{\circ}{\Pi}_{\alpha\beta}^{\gamma}=\overset{\circ}{\Gamma}%
_{\alpha\beta}^{\gamma}-(n+1)^{-1}\left[  \delta_{\alpha}^{\gamma}%
\overset{\circ}{\Gamma}_{\epsilon\beta}^{\epsilon}-\delta_{\beta}^{\gamma
}\overset{\circ}{\Gamma}_{\epsilon\alpha}^{\epsilon}\right]  ,
\label{Thomases0}%
\end{equation}

$\overset{\circ}{\Gamma}_{\alpha\beta}^{y}$ are the Christoffel symbols of
space-time $E_{4}$ whose fundamental tensor in used coordinate system is
$\eta_{\alpha\beta}$, $\Gamma_{\alpha\beta}^{\gamma}$ are the Christoffel
symbols of the Riemannian space-time $V_{4}$, whose fundamental tensor is
$g_{\alpha\beta}$. The semi-colon in eq. (\ref{myeqs}) denotes covariant
differentiation in $E_{4}$, Greek indexes run from 0 to 3.

The peculiarity of eq.(\ref{myeqs}) is that they are invariant under 
arbitrary
transformations of the tensor $g_{\alpha\beta}$ retaining invariant the
equations of motion of a test particle, i.e. geodesics lines in $V_{4}$. 
In
other words, the equations are geodesic-invariant. Thus, the tensor field
$g_{\alpha\beta}$ is defined up to geodesic mappings of space-time $V_{4}$ 
(In
the analogous way as the potential $A_{\alpha}$ in electrodynamics is
determined up to gauge transformations). A physical sense has only 
geodesic
invariant values. The simplest object of that kind is the object
$B_{\alpha\beta}^{\gamma}$ which can be named the strength tensor of
gravitation field. The coordinate system is defined by the used 
measurement
instruments and is a given.

Testing of eq.(\ref{myeqs}) by the classical effects in the solar system
\cite{Verozub2} and by the binary pulsar PSR1913+16 \cite{VerKoch2} show 
that
physical consequences from (\ref{myeqs}) do not contradict available
experimental data. They very little differ from the ones in general 
relativity
if the distance $r$ from an attractive mass $M$ is much larger than
Schwarzschild radius $r_{g}=2GM/c^{2}$ ( $G$ is the gravitational constant ).
However, they are complete different if $r$ became of the order of $r_{g}$ or
less than that since the event horizon is absent.

\section{Evolution of an Expanding Dust -Ball}

Consider in flat space-time dynamics of a self-gravitating spherically
symmetric homogeneous expending dust - ball with the mass $M$ .

The motion of the specks of dust with the masses $m_{p}$ in the 
spherically
symmetric field are described by the Lagrangian \cite{Verozub1}, 
\cite{Verozub2}
\begin{equation}
L=-m_{p}c\left[  c^{2}C-A\dot{r}^{2}-f^{2}(\dot{\varphi}^{2}\sin^{2}%
\theta+\dot{\theta}^{2})\right]  ^{1/2}, \label{lagrangian2}%
\end{equation}
where
\[
A=r^{4}/f^{4}(1-r_{g}/f),\ \ \ C=1-r_{g}/f,
\]%

\[
f=(r_{g}^{3}+r^{3})^{1/3},\ \ \ r_{g}=2GM/c^{2},
\]
$\dot{r}=dr/dt$, $\dot{\varphi}=d\varphi/dt$, $\dot{\theta}=d\theta/dt$.

The differential equation of particles radial motion of the ball surface 
is
given by (See also \cite{Verozub3}, \cite{Verozub0}).
\begin{equation}
\dot{R}^{2}=\frac{c^{2}C}{A}\left[  1-\frac{C}{\bar{E}^{2}}\right]  ,
\label{eq1_motion}
\end{equation}
where $R$ is the radius of the ball, $\dot{R}=R/dt$, 
$\bar{E}=E/m_{p}c^{2}$
and $E$ is the energy of the specks of dust, $C$ and $A$ are the functions of
$R$.

\begin{figure}[tbh]
\includegraphics[width=55mm,height=55mm]{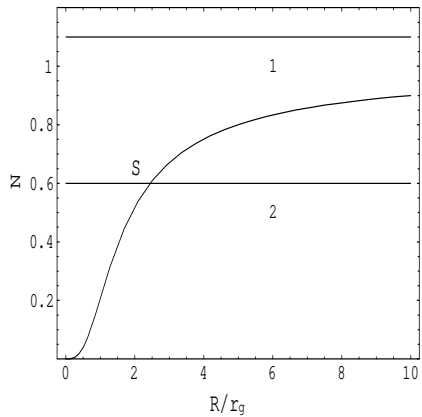}\caption{The effective
gravitational potential $\mathcal{N}(R/r_{g})$ }%
\label{NR}%
\end{figure}

Setting $\dot{R}=0$ in eq. (\ref{eq1_motion}) we obtain
\begin{equation}
\bar{E}^{2}=\mathcal{N}(R), \label{efpot}%
\end{equation}
where
\[
\mathcal{N}(R)=1-\frac{r_{g}}{f}%
\]
The function $N(R)$ is the effective potential of the spherically 
symmetric
gravitational field in the theory under consideration. Fig. \ref{NR} 
shows 
$N$
as the function of $R/r_{g}$ .

The straight lines $\bar{E}^{2}=Const$ (denoted as 1 and 2) show possible 
the
expansion scenarios :

1. The orbits with $\bar{E}^{2}\geq1$ (Line 1). The expansion begins at 
$R=0$
and continues to the infinity.

2. The orbits with $\bar{E}^{2}\leq1$ (Line 2). The expansion begins at 
$R=0$
and continues up to the point $S$ of the line crossing with the curves 
$N(R)$.

The time
\begin{equation}
T=\int_{R_{in}}^{R_{0}}\overset{.}{R}^{-1}dR\label{time_univ}%
\end{equation}
of the expansion tends to infinity if $R_{in}$ tends to zero. Therefore, 
only
an ''asymptotic'' singularity in the infinitely remote past occurs in the
model if we neglect the matter pressure.

\begin{figure}[tbh]
\includegraphics[width=55mm,height=55mm]{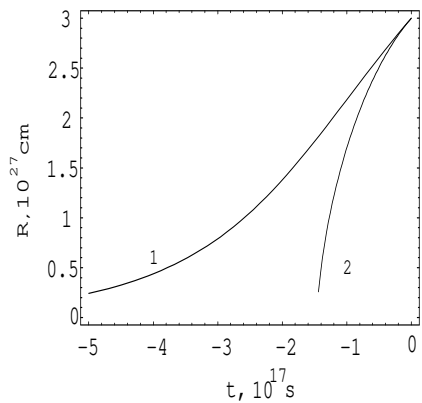}\caption{The radius of 
the
ball as the function of time.}%
\label{R_t}%
\end{figure}

For an illustrative example, let us assume that at the moment the radial
velocity
is $V_{0}=H_{0}R_{0}$ where $H_{0}=0.25\cdot10^{-17}s^{-1}$ is the Habble
constant and $R_{0}=3\cdot10^{27}cm$ is the radius. For these parameters 
Fig.
\ref{R_t} shows the function $R(t)$ (Curve 1). Curve 2 is the same 
function
for the Newtonian gravity law.

\begin{figure}[tbh]
\includegraphics[width=55mm,height=55mm]{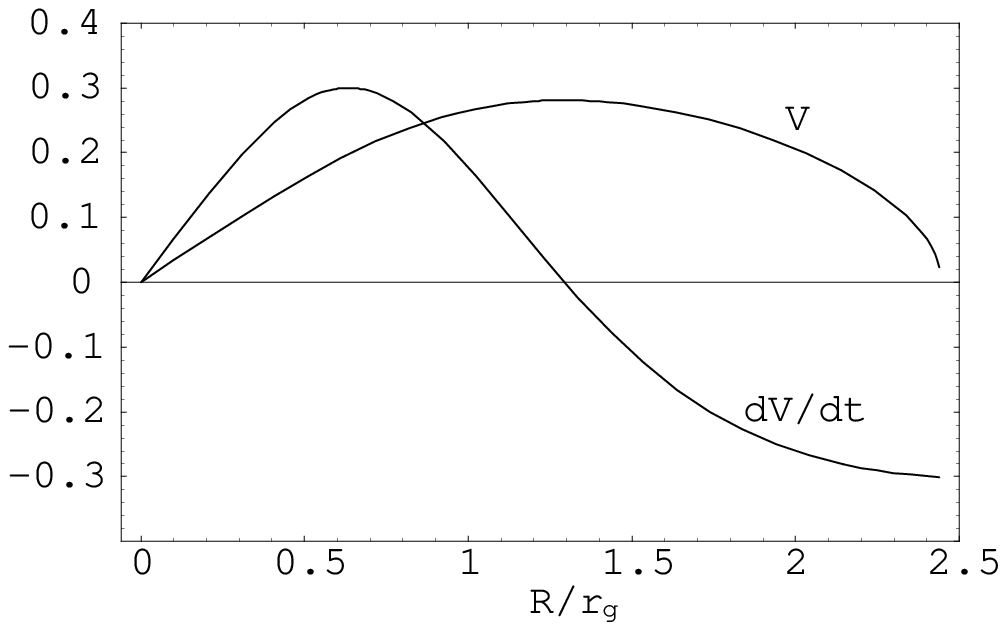}
\hfill
\includegraphics
[width=55mm,height=55mm]{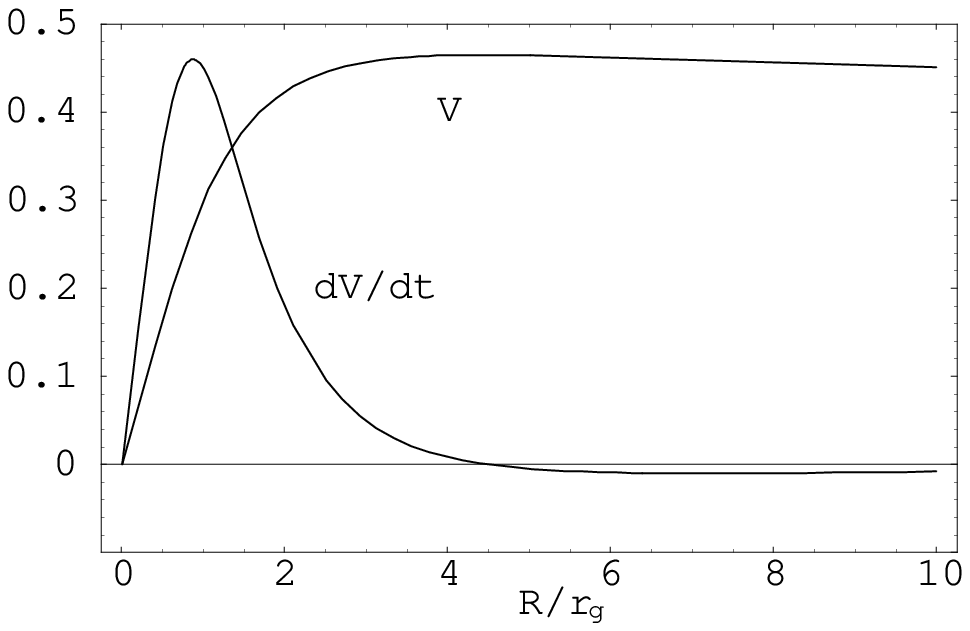} 
\newline \parbox[t]{0.47\textwidth}
{\caption
{Velocity and acceleration (in CGS units) at $\bar{E}^{2}<1$.}
\label{accel1}} 
\hfill
\parbox[t]{65mm}{\caption
{Velocity and acceleration (in CGS units) at 
$\bar{E}^{2}>1$ }\label{accel2}} 
\end{figure}

Figs. \ref{accel1} and \ref{accel2} shows the plot of the velocity
$V=\dot{R}$
and the acceleration $\overset{..}{R}=(dV/dR)V$ of the specks of dust as 
the
function of $R/r_{g}$ at $\bar{E}^{2}<1$ and $\bar{E}^{2}>1$ for the above
parameters and the density $\rho=10^{-28}$ $gm/cm^{3}$. (The values of
$dV/dt$ are muliplyed by factor 6 for convenience of comparison of the
plots).

It follows from the plot that starting from some radius $R_{c}$ the
acceleration of the selfgravitating ball become negative. This unexpected 
from
the Newtonian mechanics viewpoint fact is a consequences of the 
peculiarity 
of
the gravity force at $r\lessapprox$ $r_{g}$ \cite{Verozub1}. At 
sufficiently
large radiuses $R$ it  become close to $r_{g}$. At these distances and 
inside
the sphere $r=r_{g}$ the gravity force is repulsive. At $\bar{E}^{2}>1$ 
this
distance $R_{c}>$ $r_{g}.$ The larger $\bar{E}^{2},$ the larger is the
distance when it hapens.

\section{Dependence ''Distance - Redshift''}

We can apply the above model to a real local area of the homogeneous 
isotropic
Universe if in the theory under consideration the matter outside of the 
ball
does not create gravitational field inside the one. It is endeed take 
place
since the function $C$ in eq. (\ref{lagrangian2}) for the general  
spherically
symmetric solution is of the form \cite{Verozub1}%

\begin{equation}
C=1-\alpha/f,
\end{equation}
where $f=(\beta^{3}+r^{3})^{1/3}.$ The constant $\alpha$ is determined 
from
the correspondence principle with the nonrelativistic limit (Newtonian
theory). For this reason it must be equal to zero inside the sphere .

The luminosity distance $D_{L}$ is the following function of the redshift
$z=(\omega-\omega_{0})/\omega_{0},$ where $\omega$ and $\omega_{0}$ are
frequencies of the emitted and received light, correspondingly,
\cite{Zeldovich}:%
\begin{equation}
D_{L}=R\left(  z\right)  \left(  1+z\right)  ^{2}.
\end{equation}
where $R(z)$ is the distance to a remote galaxy with redshift parameter 
$z$.

The value $R(z)$ is a distance $r$ from the center of the ball at the 
moment
when a galaxy emited the photon that had the redsift $z$. The equation of 
the
radial motion of a photon is given by \cite{Verozub1}
\begin{equation}
\overset{.}{r}=-c\left(  \frac{C}{A}\right)  ^{1/2}%
\end{equation}
Therefore, (in an analogy with \cite{Zeldovich} ) $R(z)$ can be found by
solution of the differential equation%
\begin{equation}
\frac{dr}{dz}=-c\left(  \frac{C}{A}\right)  ^{1/2}\left(  \frac{dz}%
{dt}\right)  ^{-1},
\end{equation}
where $C$ and $A$ are the functions of $r(z)$ and the function 
$dz/dt$ of $z$
to be supposed as known.

A shift of the frequency $\omega=2\pi\nu$ at the Doppler shift of a 
remote
objects at its moving from $R$ to $R+dR$ is $d\omega=-c^{-1}H\ \omega\ dR$
which together with relation $\overset{.}{R}=H\ R$ yields
\begin{equation}
R\ \omega=Const.
\end{equation}
As a consequence of this equation and the definition of $z$ we obtain%

\begin{equation}
R=R_{0}/(1+z),\label{RZ}%
\end{equation}
where $R_{0}$ is the distance to the galaxy at the moment.

By using (\ref{RZ}) and taking into account that 
$\rho=\rho_{0}(1+z)^{3}$ ,
were $\rho_{0}$ is the presently matter density, we obtain from the 
equation
$\overset{.}{\rho}+\nabla(\rho\overset{.}{R})=0$%
\begin{equation}
\frac{dz}{dt}=-H\ (1+z).
\end{equation}

The function $H(z)$ can be found by substitution $\dot{R}=HR$ into eq.
\ref{eq1_motion} :%

\begin{equation}
H=\frac{c}{R}\left[  \frac{C}{A}\left(  1-\frac{C}{\bar{E}^{2}}\right)
\right]  ^{1/2}, \label{HubleConstantOfTime}%
\end{equation}
In this equation

\ \
\begin{equation}
\bar{E}^{2}=\frac{c^{2}f_{0}(f_{0}-r_{g})^{3}}{c^{2}f_{0}^{2}(f_{0}-r_{g}%
)^{2}-H_{0}^{2}R_{0}^{6}},\label{ConstantE}%
\end{equation}
where $f_{0}=(R_{0}^{3}+r_{g}^{3})^{1/3}$, $r_{g}=8\pi G\rho_{0}R_{0}%
^{3}/3c^{2}$,  and the equation $\rho R^{3}=\rho_{0}R_{0}^{3}$ where used.

Finally,
\begin{equation}
\frac{dR}{dz}=\frac{R}{(1+z)\left(  1-C/\overline{E}^{2}\right)  ^{1/2}},
\label{DR/DZ}%
\end{equation}
where in the constant $\overline{E}$ $R_{0}=R(z)(1+z)$. An integration of
\ref{DR/DZ} at the initial condition $R(0)=0$ yields an equation
\begin{equation}
R=R(z,H_{0},\Omega),
\end{equation}
where $\Omega=\rho_{0}/\rho_{c}$ and $\rho_{c}=3H_{0}^{2}/8\pi G.$ The
parameters $H_{0}$ and $\Omega$ are determined from observations.

\section{\bigskip Comparison with observation data}

In paper \cite{Riess} distance modulus
\begin{equation}
\mu=5\log D_{L}+25
\end{equation}
for 10 Type Ia supernovae (SNe Ia) in range $0.16$ 
$\leqslant$ z $\leqslant$
0.62 and 27 nearby supernovae with $z\leqslant$ 0.1 were presented. 
tHE value of
$\mu$ were determined by the multicolored light curve shape method (MLCS) 
and
by the template-fitting method.

The likelihood for the cosmological parameters $H_{0}$ and $\Omega$ can be
determined from a $\chi^{2}$ statistic, where%

\begin{equation}
\chi^{2}(H_{0},\Omega)=\sum_{i}\frac{\left[  \mu_{i}(z_{i},H_{0},\Omega
)-\mu_{0,i}\right]  ^{2}}{\sigma_{\mu_{0},i}^{2}+\sigma_{\nu}^{2}},
\end{equation}
$\mu_{0,i}$ and $\sigma_{\mu_{0},i}$ are distance modulus and the 
dispersion
in galaxy redshift (in units of the distance modulus), respectively. We 
use
value of $\sigma_{\nu}=200km/s$ for SNe Ia with small $z$ and 
$\sigma_{\nu
}=2500km/s$ for SNe Ia with large $z$ \cite{Riess}. We found that the 
Hubble
constant $H_{0}=65.7\pm$ $1.4\ km$\ $s^{-1}Mpc^{-1}$ by using MLCS-method 
and
$H_{0}=63.3\pm$ $1.5\ km\ s^{-1}Mpc^{-1}$ by using the template-fitting
method. For this reason, following to Riess at all \cite{Riess} 
argumentation,
we assume here that $H_{0}=65\pm$ $7\ km\ s^{-1}Mpc^{-1}.$

\begin{figure}[tbh]
\includegraphics[width=55mm,height=55mm]{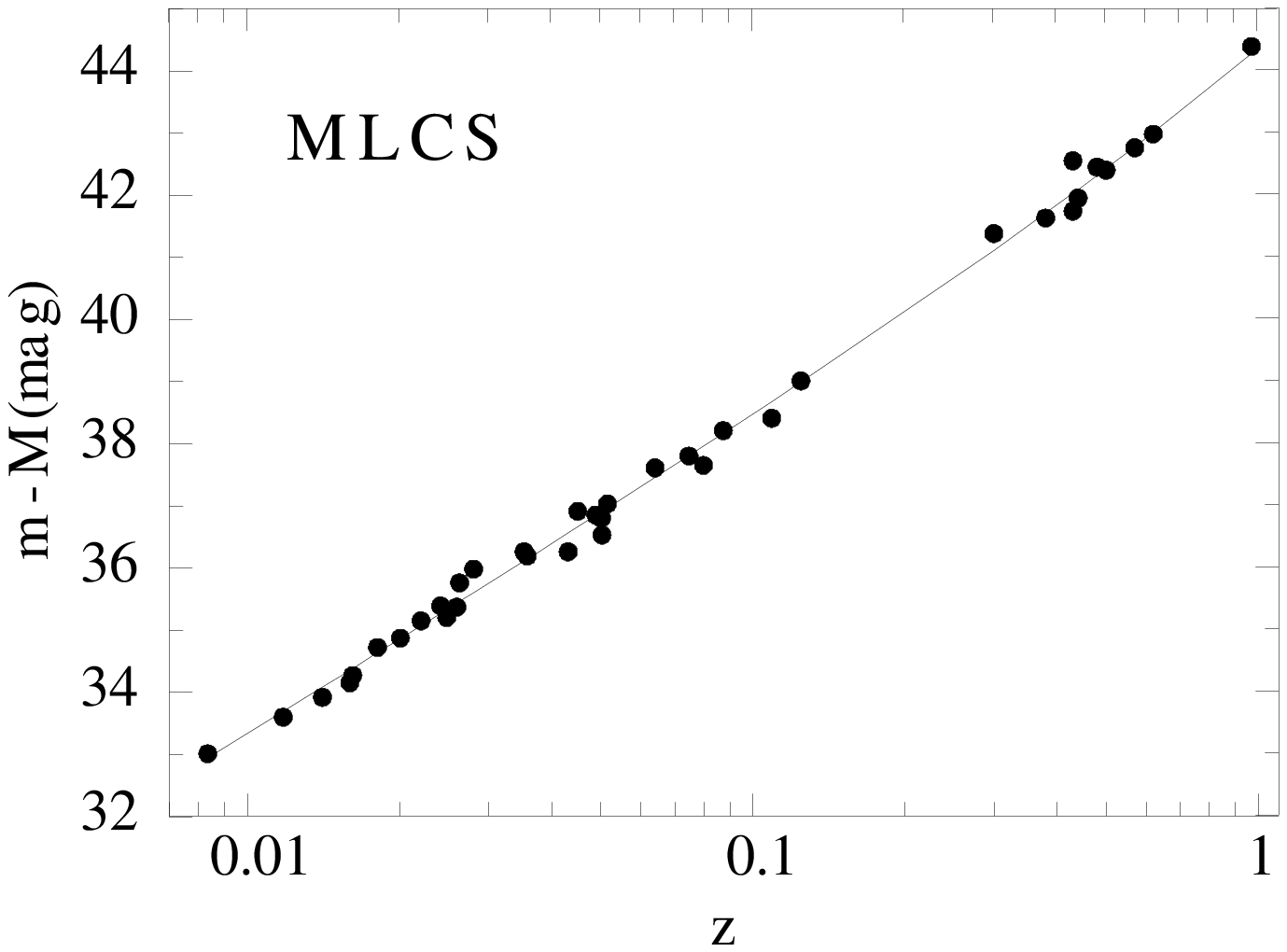} 
\hfill
\includegraphics
[width=55mm,height=55mm]{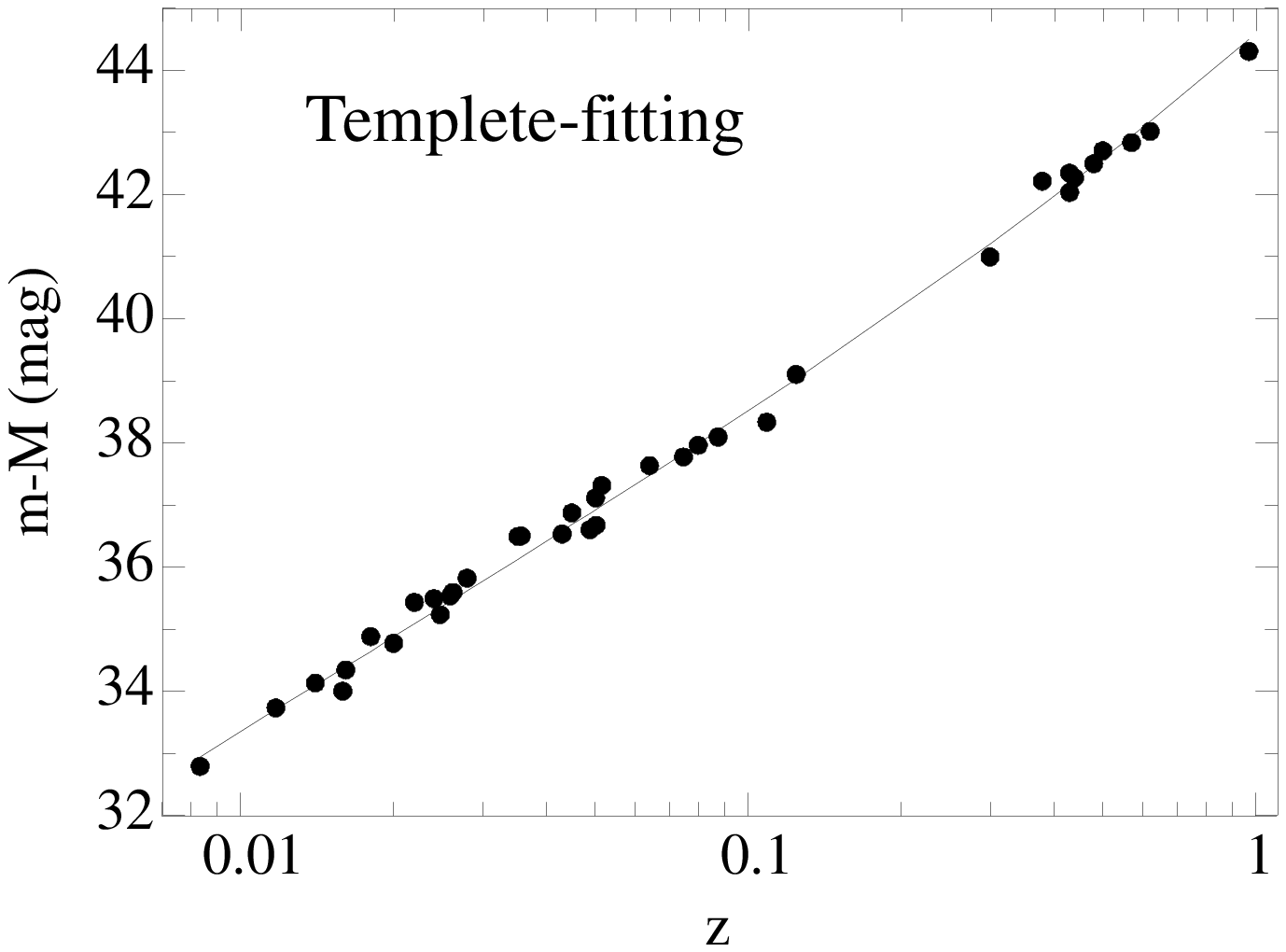}
\newline \parbox[t]{0.47\textwidth}
{\caption{The function $\mu$ of $z$ . The MLCS-method. $\Omega=0.93$ }
\label{mu1}} 
\hfill
\parbox[t]{0.47\textwidth}{\caption{The function $\mu
$ of $z$. The template-fitting method. $\Omega
= 0.39$}.\label{mu2} } 
\end{figure}

Proceeded from the data of paper \cite{Riess} we found that $\Omega
=0.93\pm0.36$ at the $93.5\%$ $\left(  1.9\sigma\right)  $ confidence 
level
for MLCS-method, and $\Omega=0.39\pm0.24$ at the $91.0\%$ $\left(
1.7\sigma\right)  $ confidence level for template-fitting method. 
(Must be
noted that the value of found parameter $\Omega$ do not depend on the 
above
found value of the Hubble constant).The function $\mu=\mu(z)$ determined 
by
the both methods are shown in Figs. \ref{mu1} and \ref{mu2} by continuous
curves. The points denote $\mu$ versus $z$ for SNe Ia from 
paper\cite{Riess}.

Using the above values of $H_{0}$ and $\Omega$ we can find the 
acceleration 
parameter%

\begin{equation}
q_{0}=-\frac{\overset{..}{R}R}{\overset{.}{R}^{2}},
\end{equation}
where $\overset{..}{R}=(d\dot{R}/dR)$ $\dot{R},$ and $\dot{R}$ is given 
by eq.
(\ref{eq1_motion}). Unlike the general relativity the acceleration 
parameter
is not a constant and according to Section 2 is a function of the distance for
galaxy or the redshift. The following equation is valid%

\begin{equation}
q_{0}(z)=R(z)\left[  \frac{C^{\prime}}{C}\frac{2C-\overline{E}^{2}%
}{2(\overline{E}^{2}-C)}+\frac{A^{\prime}}{2A}\right]  .
 \end{equation}
where the prime denotes a derivative with respect to $R$.

\begin{figure}[tbh]
\includegraphics[width=55mm,height=55mm]{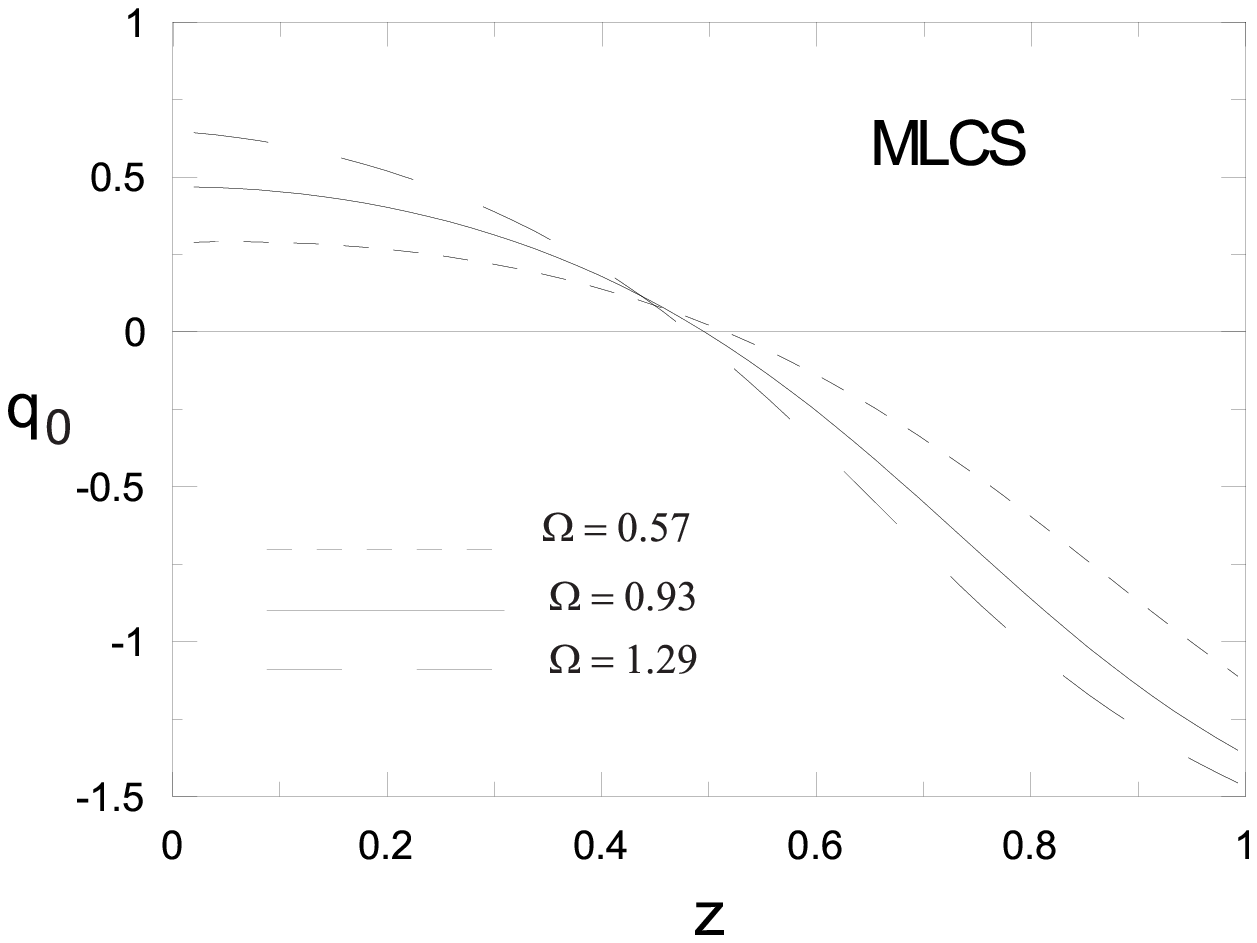} 
\hfill
\includegraphics
[width=55mm,height=55mm]{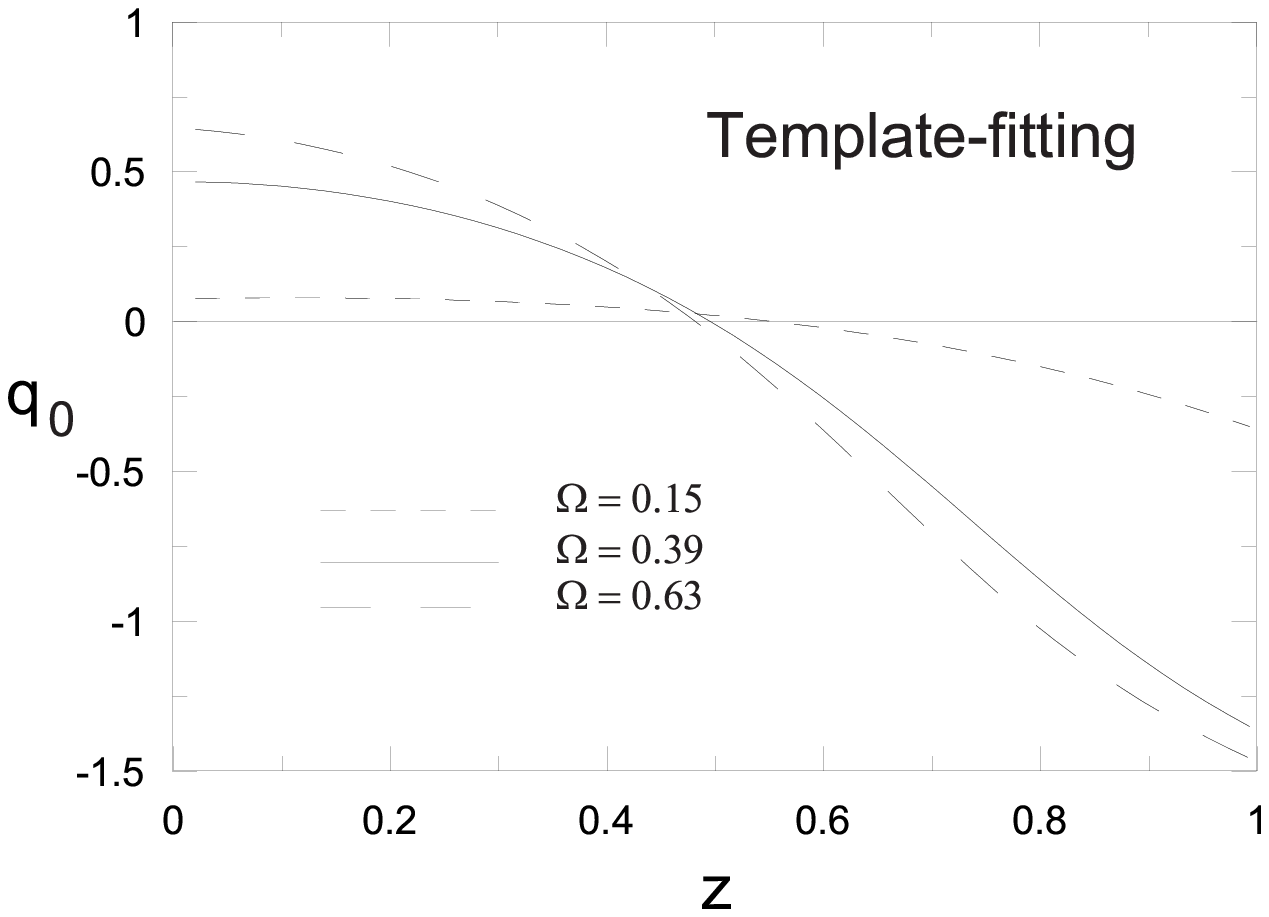}
\newline \parbox[t]{0.47\textwidth}
{\caption{The deceleration parameter found by the MLCS method 
($\Omega=0.93$)
versus $z$ (The continuous curve)
The section-line curves are the functions for $\Omega=0.93\pm0.36$ }
\label{q1}} 
\hfill
\parbox[t]{0.47\textwidth}
{\caption{The deceleration parmeter found by the template-
fitting method  ($\Omega=0.39) $ versus $z$. (The continuous curve).
The section-line curves are the
function
for $\Omega=0.39 \pm0.24$}.
\label{q2} } 
\end{figure}

Plots of the resulting function $q_{0}(z)$ for two used methods is shown 
in
Figs. \ref{q1} and \ref{q2} .(The continuous curves).

\section{Conclusion}
The recent results by two teams (the Supernova Cosmology Project and the
High-z Supernova Search Team) \cite{Riess}, \cite{Perlmutter} leads to
fundamental problems. Severeal problems are rewied by 
S. Weinberg \cite{Weinberg}. The above a simple model show that these 
results
can be also interpreted as evidence for non-Newtonian law of gravitation
proposed in \cite{Verozub1}, \cite{Verozub2}.

\end{document}